\documentstyle[12pt]{article}
\newcommand{\ba}{\begin{eqnarray}}  
\newcommand{\ea}{\end{eqnarray}}    
\newcommand{\be}{\begin{equation}}   
\newcommand{\ee}{\end{equation}}     

\newcommand{\R}{{\check{R}}^{(\frac{1}{2}, \frac{1}{2})}}     
\newcommand{\Ra}{\check{R}}
\newcommand{\Rb}{{\check{R}}^{(1,1)}}
\newcommand{\Rc}{\check{\cal R}}

\newcommand{\ga}{\gamma}

\newcommand{\dl}{\delta}

\newcommand{\la}{\lambda}

\newcommand{\sh}{\sinh}
\newcommand{\ch}{\cosh}
\newcommand{\ol}{\overline}

\def\frac#1#2{{#1\over #2}}
\def\dfrac#1#2{{\displaystyle{#1\over#2}}}

\begin{document}

\centerline{\Large \bf Quantum Dynamical $\Ra$- Matrix  } 
\vspace{2mm}
\centerline{\Large \bf with Spectral Parameter from Fusion }
\vspace{5mm}
\centerline{\large Xu-Dong Luo$^{\dag}$, Xing-Chang Song$^{\ddag\dag}$, Shi-Kun Wang$^{\S}$ and Ke Wu$^{\dag}$}
\vspace{3mm}
\begin{minipage}{15cm}
\ddag \quad Department of Physics, Peking University, Beijing 100871, China 

\dag \quad Institute of Theoretical Physics,  Academia Sinica,  Beijing 100080, China  

\S \quad CCAST(World Laboratory), $\;\;$ Beijing 100080, China  
 
\quad and  $\;$ Institute of Applied Mathematics,  Academia Sinica,  Beijing 100080, China 

\vspace{9mm}

\centerline{\bf Abstract}
\vspace{5mm}

\quad A quantum dynamical $\Ra$-matrix with spectral parameter is constructed by fusion procedure.
 This spin-1 $\Ra$-matrix is connected with Lie algebra $so(3)$  and does not satisfy the condition of translation invariance.    
\end{minipage}

\section{Introduction}
\indent 
 Since the classical dynamical $r$-matrix \cite{Avan-Talon} first appeared on the scene of integrable many body system, 
many dynamical $r$-matrices have been found in integrable models such as Calogero-Moser model \cite{Sklyanin}, 
Sine-Gorden soliton case \cite{Babelon} and the general case for the Ruijsenaars systems \cite{Avan-Rollet}. These dynamical 
$r$-matrices do not satisfy the ordinary classical Yang-Baxter equation, so its quantization is rather nontrivial. 
The quantum dynamical Yang-Baxter (QDYB) equation, which appeared first in the quantization of Toda field theory \cite{Gervais} 
and later in the quantization of KZB equation \cite{Felder}, had been studied widely for various integrable models and its algebraic 
structure was explored \cite{{Avan},{Arnaudon}, {Etingof}}.  

\indent
In contrast to the non-dynamical one \cite{Ma}, only a few dynamical $R$ matrices are constructed explicitly and most of them can 
be obtained from Felder's
solution \cite{Felder} by taking a gauge transformations \cite{Etingof}. So how to construct new $R$ matrix is still an 
interesting and challenging problem. As an efficient method to obtain higher-spin $R$ matrix, fusion procedure \cite{Kulish}
has been applied to dynamical $R$ matrix \cite{Song}. 

\indent In this paper, we construct a spin-1 quantum dynamical $R$-matrix with 
spectral parameter by "fusing" together the spin- $\frac{1}{2}$ $R$-matrices which satisfy the 
QDYB equation \cite{Avan}:
\be
\begin{array}{c}
  R_{12}(\la_{12},x+\ga h^{(3)})R_{13}(\la_{13},x-\ga h^{(2)})R_{23}(\la_{23},x+\ga h^{(1)})  \\ 
= R_{23}(\la_{23},x-\ga h^{(1)})R_{13}(\la_{13},x+\ga h^{(2)})R_{12}(\la_{12},x-\ga h^{(3)}).
\end{array}
\ee 
Where the spectral parameters $\la_{ij}$ are defined as $\la_{ij}=\la_{i}-\la_{j} $, 
$x=\sum\limits_{\nu}x_{\nu}h_{\nu}$ is the dynamical variable and $h$ is the Cartan subalgebra of the
underlying simple Lie algebra. Taking values in End$(V_{1}\otimes V_{2}\otimes V_{3})$, $R$ matrix appears as 
$R_{12}(x+\ga h^{(3)})(V_{1}\otimes V_{2}\otimes V_{3})=(R_{12}(x+\ga \mu)(V_{1}\otimes V_{2}))\otimes V_{3}$ 
if $h^{(3)}$ has weight $\mu$ in space $V_{3}$. Other symbols have a similar meaning.   

\indent In braid form, the QDYB equation (1) reads as
\be
\begin{array}{c}
\Ra_{23}(\la_{12},x+\ga h^{(1)})\Ra_{12}(\la_{13},x-\ga h^{(3)})\Ra_{23}(\la_{23},x+\ga h^{(1)})  \\
=\Ra_{12}(\la_{23},x-\ga h^{(3)})\Ra_{23}(\la_{13},x+\ga h^{(1)})\Ra_{12}(\la_{12},x-\ga h^{(3)}),
\end{array}
\ee 
where $\Ra_{ij}=P_{ij}R_{ij}$ and $P_{ij}$ is the permutation operator acting on spaces
$V_{i}\otimes V_{j}$. If $\Ra$ matrices satisfy the condition of translation invariance:
\be
\left[ {\cal D}^{(i)}+{\cal D}^{(j)} \; , \; \check{R}_{ij}(\la,x) \right] =0  \; ; 
 \quad {\cal D}^{(i)}=\sum\limits_{\nu}h^{(i)}_{\nu} \partial_{x_{\nu}} ,  
\ee 
 we can rewrite equation (2) as 
\be
\begin{array}{l}
 \; \Ra_{23}(\la_{12},x+2\ga h^{(1)})\Ra_{12}(\la_{13},x)\Ra_{23}(\la_{23},x+2\ga h^{(1)})  \\
=\Ra_{12}(\la_{23},x)\Ra_{23}(\la_{13},x+2\ga h^{(1)})\Ra_{12}(\la_{12},x) .
 \end{array}
\ee
\indent
This paper is organized as follows. In section 2, we obtain some useful properties of $\R$ matrix.
In section 3, using the $\R$ matrix, we construct the $\Rb$ matrix by fusion procedure and prove that the new matrix satisfies the QDYB 
equation too. Finally, we discuss our results and compare it with paper \cite{Song} in section 4.

\section{Properties of spin- $\frac{1}{2}$ $\Ra$-matrix}  
\indent 
 According to spin- $\frac{1}{2}$ chain, $h^{(i)}(\otimes V_{i})=diag\{ \frac{1}{2},-\frac{1}{2}\} (\otimes V_{i})$, 
there is the simplest $\check{R}$ matrix solution with spectral parameter \cite{Avan}:
\be
\R(\la,x)=\left(\begin{array}{cccc}
1   &0   &0  &0\\
0   &-\dfrac{\sh\ga \sh(x+\la)}{\sh x \sh(\la-\ga)}   & \dfrac{\sh\la \sh(x+\ga)}{\sh x \sh(\la-\ga)}  &0\\
0   &\dfrac{\sh\la \sh(x-\ga)}{\sh x \sh(\la-\ga)}  & -\dfrac{\sh\ga \sh(x-\la)}{\sh x \sh(\la-\ga)}  &0\\
0   &0   &0  &1
\end{array}
\right).
\ee
This $\R$ matrix satisfies the "weight zero" condition
\be
\left[ h^{(i)}+h^{(j)} \; , \; \check{R}_{ij}(\la,x) \right] =0 ,  
\ee
and it has one triple eigenvalue $1$ and one single eigenvalue $-\dfrac{\sh(\la+\ga)}{\sh(\la-\ga)}$.

\indent To the triple eigenvalue, its right-acting eigenvectors are
\ba
u_{(1)}(x)=\left( \begin{array}{c}
       1 \\ 0 \\  0 \\  0 
       \end{array} \right) \; ;   &
u_{(0)}(x)=\dfrac{1}{\sqrt{2}}\left( \begin{array}{c}
       0 \\ 1 \\ 1 \\ 0 
        \end{array} \right)\; ;    &
u_{(-1)}(x)=\left( \begin{array}{c}
       0 \\ 0 \\ 0 \\ 1 
        \end{array} \right),
\ea 
and its left-acting eigenvectors are
\be
\begin{array}{ccl}
{\ol{u}}^{(1)}(x) &=& (1,\; 0,\; 0,\; 0)  \\
{\ol{u}}^{(0)}(x) &=& \dfrac{1}{\sqrt{2}}(0,\; \dfrac{\sh(x-\ga)}{\sh x \ch\ga},\; \dfrac{\sh(x+\ga)}{\sh x \ch\ga},\; 0)  \\
{\ol{u}}^{(-1)}(x) &=& (0,\; 0,\; 0,\; 1).  
\end{array}
\ee
While the eigenvalue is $-\dfrac{\sh(\la+\ga)}{\sh(\la-\ga)}$, the right-acting 
and left-acting eigenvectors are
\ba
v_{(0)}(x)=\dfrac{1}{\sqrt{2}}\left( \begin{array}{c}
       0 \\ 
       \dfrac{\sh(x+\ga)}{\sh x \ch\ga} \\ 
       \dfrac{-\sh(x-\ga)}{\sh x \ch\ga}  \\ 
        0 
       \end{array} \right) \; ;   &
{\ol{v}}^{(0)}(x) =\dfrac{1}{\sqrt{2}} (0,\; 1,\; -1,\; 0)        
\ea
respectively. 

\indent  These eigenvectors satisfy 
\be
\begin{array}{cc}
{\ol{u}}^{(a)}(x) v_{(0)}(x)={\ol{v}}^{(0)}(x) u_{(a)}(x)=0 \;\; , \quad & a=1,0,-1    \\
{\ol{v}}^{(0)}(x) v_{(0)}(x)=1 \; ; \quad  {\ol{u}}^{(a)}(x) u_{(b)}(x)=\dl^{a}_{\; b} \; ,\quad & \quad a, \; b=1,0,-1  
\end{array}
\ee
so we can construct two projection operators for the triplet and singlet
\be
\begin{array}{l}
P(x) = \sum\limits_{a} u_{(a)}(x){\ol{u}}^{(a)}(x)  ; \quad Q(x) = v_{(0)}(x){\ol{v}}^{(0)}(x)  \\
id_{(4 \times 4)}=  P(x) + Q(x),
\end{array}
\ee
in which $id_{(4 \times 4)}=diag\{ 1,1,1,1\}$, $P(x)$ and $Q(x)$ have the properties: 
 $$P^{2}(x)=P(x)\; ; \quad Q^{2}(x)=Q(x)\; ; \quad P(x)Q(x)=Q(x)P(x)=0$$
 $$P(x)u_{(a)}(x)=u_{(a)}(x)\; , \quad {\ol{u}}^{(a)}(x) P(x)={\ol{u}}^{(a)}(x)\; ; \quad a=1,0,-1 .$$
Now, we can rewrite $\R(\la,x)$ as
$$
\R(\la,x)=P(x) -\dfrac{\sh(\la+\ga)}{\sh(\la-\ga)} Q(x). 
$$
\indent It is obvious that 
\be
\R(\la=-\ga,x)=P(x).
\ee  
Applying this property to equation (2), we obtain
\be
\begin{array}{c}
  P_{23}(x+\ga h^{(1)}) \R_{12}(\la-\ga,x-\ga h^{(3)}) 
\R_{23}(\la,x+\ga h^{(1)})    \\
 = \R_{12}(\la,x-\ga h^{(3)}) \R_{23}(\la-\ga,x+\ga h^{(1)}) 
P_{12}(x-\ga h^{(3)})   \\ \\
 \R_{23}(\la,x+\ga h^{(1)}) \R_{12}(\la-\ga,x-\ga h^{(3)}) 
P_{23}(x+\ga h^{(1)})    \\
 = P_{12}(x-\ga h^{(3)}) \R_{23}(\la-\ga,x+\ga h^{(1)}) 
\R_{12}(\la,x-\ga h^{(3)}).
\end{array}
\ee 
\section{Construct spin- $1$ $\Ra$-matrix}
\indent  
Refer to fusion procedures in papers \cite{{Kulish},{Song}},  
we "fuse" dynamical $\Rb$ matrix with spectral parameter as follows  
\be
\begin{array}{l}
 \left[ \Rb_{12,34}(\la,x) \right]^{ab}_{cd} =   \\
      \quad  {\ol{u}}^{(a)}_{12}(x-\ga h^{(3,4)}) 
           {\ol{u}}^{(b)}_{34}(x+\ga h^{(1,2)})
          \R_{23}(\la+\ga,x+\ga h^{(1)}-\ga h^{(4)})
          \R_{12}(\la,x-\ga h^{(3,4)})   \\
      \quad \times  \R_{34}(\la,x+\ga h^{(1,2)}) 
                   \R_{23}(\la-\ga,x+\ga h^{(1)}-\ga h^{(4)})
                   {{u}_{12}}_{(c)}(x-\ga h^{(3,4)})
                   {{u}_{34}}_{(d)}(x+\ga h^{(1,2)})
\end{array}
\ee
in which $a,b,c,d$ take values among $1,0,-1$ and $h^{(i,j)}$ means $h^{(i)}+h^{(j)}$, so this $\Rb$ matrix is a $9\times 9$ matrix. 

\indent In order to prove that equation (14) satisfies QDYB equation too, we define two $4 \times 4$ matrices as follows:
\begin{eqnarray*}
u=(u_{(1)},u_{(0)},0,u_{(-1)}) \; ;  \quad \ol{u}=\left( \begin{array}{c}
       \ol{u}^{(1)} \\ 
       \ol{u}^{(0)}  \\ 
         0     \\ 
       \ol{u}^{(-1)}   
       \end{array} \right).   
\end{eqnarray*}        
then, we replace ${\ol{u}}^{(a)}$ and ${\ol{u}}^{(b)}$ by $\ol{u}$ as well as replacing ${u}_{(c)}$ and ${u}_{(d)}$ by $u$ 
in equation (14),
such that $\Rb$ is changed into a $16\times 16$ matrix, where the added seven rows and seven columns
are nothing but zero in fact. Such $u$ and $\ol{u}$ matrices not only keep $u(x)\ol{u}(x)=P(x)$, $P(x)u(x)=u(x)$ and
$\ol{u}(x) P(x)=\ol{u}(x)$, but also satisfy the weight zero condition too. Now the QDYB equation becomes
\be
\begin{array}{c}
\Rb_{34,56}(\la_{12},x+\ga h^{(1,2)})\Rb_{12,34}(\la_{13},x-\ga h^{(5,6)})\Rb_{34,56}(\la_{23},x+\ga h^{(1,2)})  \\ \\
=\Rb_{12,34}(\la_{23},x-\ga h^{(5,6)})\Rb_{34,56}(\la_{13},x+\ga h^{(1,2)})\Rb_{12,34}(\la_{12},x-\ga h^{(5,6)}).
\end{array}
\ee
For simplicity, we introduce $\Rc_{ij}(\la) := \R_{ij}(\la,x+\ga \sum\limits_{k=1}^{i-1} h^{(k)}-\ga \sum\limits_{l=j+1}^{6} h^{(l)})$,
and replace $u_{ij}(x+\ga \sum\limits_{k=1}^{i-1} h^{(k)}-\ga \sum\limits_{l=j+1}^{6} h^{(l)})$ and
$\ol{u}_{ij}(x+\ga \sum\limits_{k=1}^{i-1} h^{(k)}-\ga \sum\limits_{l=j+1}^{6} h^{(l)})$ 
by ${\cal u}_{ij}$ and ${\cal \ol{u}}_{ij}$ respectively.
After these notations, the weight zero condition means
\be
\left[ A_{i\; i+1}(\la),\; B_{j\; j+1}(\la^{\prime}) \right]=0 ;\quad \; if \quad i+1<j \;\; or \;\; j+1<i 
\ee
in which $A,B \in \{\Rc,{\cal u},{\cal \ol{u}}\} $. By the relation (13) and its analogue, we can reduce equation (15) to
\begin{eqnarray*}
   l.h.s. &= & {\cal \ol{u}}_{12} {\cal \ol{u}}_{34} {\cal \ol{u}}_{56} 
   S_{34}(\la_{12}) S_{12}(\la_{13}) S_{34}(\la_{23})
    {\cal u}_{12} {\cal u}_{34} {\cal u}_{56}  \\ 
   r.h.s. &=& {\cal \ol{u}}_{12} {\cal \ol{u}}_{34} {\cal \ol{u}}_{56}
    S_{12}(\la_{23}) S_{34}(\la_{13}) S_{12}(\la_{12})
    {\cal u}_{12} {\cal u}_{34} {\cal u}_{56}   \\
   S_{i\; i+1}(\la) &=& (\Rc_{i+1\; i+2}(\la-\ga) \Rc_{i\; i+1}(\la) \Rc_{i+2\; i+3}(\la) \Rc_{i+1\; i+2}(\la+\ga)).
\end{eqnarray*}
Using QDYB equation (2) and its analogue, we have proved $S_{34}(\la_{12}) S_{12}(\la_{13}) S_{34}(\la_{23}) 
=S_{12}(\la_{23}) S_{34}(\la_{13}) S_{12}(\la_{12})$, or $l.h.s.=r.h.s.$ in above equation. In other words, the
fusion procedure is practicable. 

\indent If we rewrite equation (15) in the standard $9\times 9$ matrix form $\Rb_{IJ}(\la,x)$, it becomes
\be
\begin{array}{c}
\Rb_{JK}(\la_{12},x+\ga h^{(I)}) \Rb_{IJ}(\la_{13},x-\ga  h^{(K)}) \Rb_{JK}(\la_{23},x+\ga h^{(I)})  \\ \\
=\Rb_{IJ}(\la_{23},x-\ga h^{(K)}) \Rb_{JK}(\la_{13},x+\ga h^{(I)}) \Rb_{IJ}(\la_{12},x-\ga h^{(K)}).
\end{array}
\ee
It is just the original QDYB equation (2). Notice that this $\Rb$ matrix is of 
spin- 1 since 
$h^{(l)}(\otimes V_{l}) \; (\hbox{in which} \; l \in \{ I,J,K \} )$ becomes $diag\{ 1,0,-1\} (\otimes V_{l})  $
 by taking the singlet of spin- 0 away. 

\indent With the $\R$ matrix (5) and the fusion method (14), we obtain
\be
\begin{array}{l}
\Rb(\la,x)=         \\ \\
 \;\; \left( \begin{array}{ccc|ccc|ccc}
    1 & 0        & 0        & 0        & 0        & 0        & 0        & 0        & 0 \\  
    0 & a(\la,x) & 0        & b(\la,-x)& 0        & 0        & 0        & 0        & 0 \\  
    0 & 0        & c(\la,x) & 0        & d(\la,x) & 0        & e(\la,x) & 0        & 0 \\  \hline
    0 & b(\la,x)& 0         & a(\la,-x)& 0        & 0        & 0        & 0        & 0 \\ 
    0 & 0        & f(\la,x) & 0        & g(\la,x) & 0        & f(\la,-x)& 0        & 0 \\  
    0 & 0        & 0        & 0        & 0        & a(\la,x) & 0        & b(\la,-x)& 0 \\  \hline
    0 & 0        & e(\la,-x)& 0        & d(\la,-x)& 0        & c(\la,-x)& 0        & 0 \\  
    0 & 0        & 0        & 0        & 0        & b(\la,x) & 0        & a(\la,-x)& 0 \\
    0 & 0        & 0        & 0        & 0        & 0        & 0        & 0        & 1
      \end{array}   \right),
\end{array}
\ee
in which 
\begin{eqnarray*}
a(\la,x)=\dfrac{\sh(2\ga) \sh(\la+x)}{\sh(2\ga-\la) \sh x}  \quad\quad 
b(\la,x)=\dfrac{\sh(\la) \sh(2\ga-x)}{\sh(2\ga-\la) \sh x}  
\end{eqnarray*}
\begin{eqnarray*}
c(\la,x)&=&\dfrac{\sh\ga \sh(2\ga) \sh(\la+x) \sh(\ga+\la+x)}{\sh(\ga-\la) \sh(2\ga-\la) \sh x \sh(\ga+x)}  \\  \\
d(\la,x)&=&\dfrac{\sh(2\ga) \sh(\la) \sh(2\ga+x) \sh(\la+x) \ch\ga}{\sh(\ga-\la) \sh(2\ga-\la) \sh(\ga-x) \sh(\ga+x)}   \\ \\
e(\la,x)&=&-\dfrac{\sh\la \sh(\ga+\la) \sh(\ga+x) \sh(2\ga+x)}{\sh(\ga-\la) \sh(2\ga-\la) \sh(\ga-x) \sh x}  \\  \\
f(\la,x)&=&\dfrac{2 \sh\ga \sh\la \sh(\ga-x) \sh(\la+x)}{\sh(\ga-\la) \sh(2\ga-\la) \sh x \sh(\ga+x)}  \\ \\
g(\la,x)&=&\dfrac{\sh(\ga+\la)}{\sh(\ga-\la)}+\dfrac{\sh\la (\ch(2 x) -  \ch(2 \ga) - {\sh}^{2}(2 \ga) )}{\sh(2 \ga-\la) \sh(\ga-x) \sh(\ga+x)} .
\end{eqnarray*} 
\indent
The obtained $\Rb$ matrix has three distinct eigenvalues, say, $1$, $-\dfrac{\sh(\la+ 2\ga)}{\sh(\la- 2\ga)}$ and 
$\dfrac{\sh(\la+ \ga) \sh(\la+ 2\ga)}{\sh(\la- \ga) \sh(\la- 2\ga)}$ whose multiplicity are $5$, $3$ and
$1$ respectively. This $\Rb$ is connected with Lie algebra $so(3)$. By direct calculation, we can show that 
it does satisfy the QDYB equation (17)
with $h^{(l)}(\otimes V_{l})=diag\{ 1,0,-1\} (\otimes V_{l})$.
 
\section{Discussion}
\indent 
From the expression (18), we find the $\Rb$ matrix does not satisfy the translation invariance condition (3). 
In other words, if we want to translate 
it to the form of equation (4), we will obtain a more complex $\Rb$ matrix form. In fact, it is just the matrix of  
${{\Ra}_{IJ}^{(1,1)}}(\la,x+\ga h^{(I,J)})$ where $h^{(I,J)}$ means $h^{(I)}+h^{(J)}$, so we have to construct new commuting operators
different from those in paper \cite{Avan}, in which  the condition $(3)$ was used in constructing commuting operators. For the simplicity of 
the expression of $R$- matrix, we had better use more symmetric form as equation $(1)$ or $(2)$, rather than the form as equation (4).

\indent 
Now we compare our results with paper \cite{Song}. At first, QDYB equation (4) tends to be independent from the spectral parameter by 
requiring $\la \rightarrow \pm\infty$. 
Secondly we need change dynamical variable $x \rightarrow -\ga x$ in our $\R$ and $\Rb$ 
matrices because QDYB equation takes different forms in these two papers. At last, we need translate expression (18) 
to ${{\Ra}_{IJ}^{(1,1)}}(\la,x+\ga h^{(I,J)})$ as discussed before. 
After these changes of $\la$ and $x$ in ${{\Ra}_{IJ}^{(1,1)}}(\la,x+\ga h^{(I,J)})$, we indeed obtain the  
$\Rb$ matrix gauge equivalent to  the one in paper \cite{Song}. 
The single eigenvalue $q$ of $\R$ matrix in paper \cite{Song} is connected to $e^{\pm 2\ga}$ when we take 
$\la \rightarrow \pm\infty$ respectively.
 
\indent For the six-vertex elliptic solution of QDYB equation, eigenvalues of $\R$ matrix are not made
of one triplet and one singlet. It is still an open problem about how to construct its higher-spin $\Ra$ matrix .    

\section*{Acknowledgments}

One of authors (X.D. Luo) would like to thank C. Xiong for useful discussions.
This work was supported by Scientific Project of Department of Science and Technology in China, 
Natural
Scientific Foundation of Chinese Academy of Sciences,
Doctoral Programme Foundation Institution of Higher Education 
and Foundation of NSF.

\end{document}